\documentclass[letterpaper,prl,twocolumn,showpacs]{revtex4}
\usepackage{physics}
\usepackage{amsmath,amssymb}
\usepackage{epsfig}
\usepackage[geometry]{ifsym}
\usepackage{color}

\newcommand {\ra}{\rightarrow}

\newcommand {\pt}{\partial}
\newcommand{\I}{\mathrm i}
\newcommand{\myRe}{\mathrm{Re}}
\newcommand{\myIm}{\mathrm{Im}}
\newcommand{\myArg}{\mathrm{Arg}}
\newcommand{\mydet}{\mathrm{det}}
\newcommand{\mysech}{\mathrm{sech}}
\newcommand{\E}{\mathrm e}

\begin{document}
\title{Formation of rogue waves from the locally perturbed condensate}
\date{\today}
\author{A.\,A.~Gelash$^{1,2}$}\email{agelash@gmail.com;  gelash@srd.nsu.ru}
\affiliation{$^{1}$Novosibirsk State University, Novosibirsk, 630090, Russia}
\affiliation{$^{2}$Institute of Thermophysics, SB RAS, Novosibirsk, 630090, Russia}
\begin{abstract}
The one-dimensional focusing nonlinear Schr\"{o}dinger equation (NLSE) on an unstable condensate background is the fundamental physical model, that can be applied to study the development of modulation instability (MI) and formation of rogue waves. The complete integrability of the NLSE via inverse scattering transform enables the decomposition of the initial conditions into elementary nonlinear modes: breathers and continuous spectrum waves. The small localized condensate perturbations (SLCP) that grow as a result of MI have been of fundamental interest in nonlinear physics for many years. Here, we demonstrate that Kuznetsov-Ma and superregular NLSE breathers play the key role in the dynamics of a wide class of SLCP. During the nonlinear stage of MI development, collisions of these breathers lead to the formation of rogue waves. We present new scenarios of rogue wave formation for randomly distributed breathers as well as for artificially prepared initial conditions. For the latter case, we present an analytical description based on the exact expressions found for the space-phase shifts that breathers acquire after collisions with each other.  Finally, the presence of Kuznetsov-Ma and superregular breathers in arbitrary-type condensate perturbations is demonstrated by solving the Zakharov-Shabat eigenvalue problem with high numerical accuracy.
\end{abstract}
\maketitle
\subsection{Introduction}

The formation of extreme-amplitude waves is among the most remarkable phenomenon in the physics of wave processes. In the linear case, these events may appear only as a result of simple wave interference, whereas the interactions of nonlinear waves exhibit a wide range of nontrivial mechanisms, such as the development of modulation instability (MI) and nonlinear wave focusing~\cite{onorato2013rogue,pelinovsky2008book}. The localized extreme-amplitude events, so-called \textit{rogue waves}, are of special interest as they are observed more frequently than predicted by Gaussian statistics and can appear from relatively weak perturbations of a calm background~\cite{onorato2013rogue,akhmediev2009extreme}. This phenomenon being studied first in oceanography has been observed experimentally in different nonlinear media, such as optical fiber with Kerr nonlinearity, Bose-Einstein condensate, surface of a fluid and plasmas that demonstrates its universal nature~\cite{pelinovsky2008book,solli2007optical,efimov2010rogue,chabchoub2011rogue,moslem2011dust}.

The one-dimensional focusing nonlinear Schr\"{o}dinger equation (NLSE):
\begin{equation}\label{NLSE}
\I \psi_{t} + \frac{1}{2}\psi_{xx} + (|\psi|^{2} - A^2)\psi=0\,,
\end{equation}
is the fundamental mathematical model describing weakly nonlinear wave propagation. Here, $\psi(t,x)$ is the complex-valued envelope of the physical wavefield, $t$ and $x$ are the time and space coordinates. V.E.~Zakharov and A.B.~Shabat found that the NLSE can be completely integrated using the inverse scattering transform (IST)~\cite{zakharov1972exact}. Here we study solutions of the NLSE~(\ref{NLSE}) on the so-called \textit{condensate} background -- a simple quasi-monochromatic plane wave. The condensate solution of the equation~(\ref{NLSE}) is $\psi_0(t,x) = A$, where $A$ is the background amplitude, which we assume to be real without loss of generality. The condensate is unstable with respect to long-wave perturbations (MI phenomena, see e.g.~\cite{zakharov2009modulation}) with the following growth rate:
\begin{equation}
\Gamma(k)=k\sqrt{A^2-k^2/4}\,,
\label{MI increment}
\end{equation}
where $k$ is the perturbation wave number. In the region $0<k<2$, the amplitude of these perturbations grows at the initial stage  as $\sim e^{\Gamma t}$ in the initial (linear) stage. The nonlinear stage of MI is of fundamental interest and may lead to the formation of rogue waves~\cite{onorato2013rogue,pelinovsky2008book}.

The NLSE describes only the first order nonlinear effects. However its universality and integrability allows to capture the fundamentally important features of MI and to find analytical rogue wave solutions. Indeed, the IST links the initial NLSE wavefield with the so-called \textit{scattering data}, which play the role of elementary nonlinear modes, similar to Fourier harmonics in linear wave theory. In the case of spatially localized NLSE solutions, the scattering data are represented by the discrete (solitons) and continuous (nonlinear dispersive waves) parts of the eigenvalue spectrum of the Zakharov-Shabat auxiliary linear system (ZH system).

The IST for the spatially localized wavefield and zero background ($A=0$) was developed in~\cite{zakharov1972exact}, where the $N$-soliton solutions were found analytically and the general Cauchy problem was solved implicitly via the integral Gelfand-Levitan-Marchenko equations (GLME). In 1977, E.A.~Kuznetsov~\cite{kuznetsov1977solitons} and later Y.C.~Ma~\cite{ma1979perturbed} generalized this theory to the case of the condensate backgroung. In this model the discrete spectrum solutions, interacting with condensate, transform from solitons to the oscillating structures -- breathers. The family of NLSE breathers includes the well known solutions of Peregrine~\cite{peregrine1983water}, Kuznetsov-Ma~\cite{kuznetsov1977solitons,ma1979perturbed} and Akhmediev~\cite{akhmediev1985generation}.

The nontrivial interactions of different scattering data modes makes the system evolution complicated and produces many fundamental problems. One question is the long-term evolution of unstable small localized condensate perturbations (SLCP). The SLCP have been extensively experimentally studied in nonlinear optics and hydrodinamics for the last years (see e.g.~\cite{kibler2012observation,chabchoub2011rogue,kibler2015superregular}). In the mentioned paper~\cite{kuznetsov1977solitons}, E.A.~Kuznetsov suggested that the MI of SLCP is driven by the unstable part of continuous spectrum waves and leads to the local (i.e., in the area of the initial perturbation) Fermi-Pasta-Ulam recurrence (see also his work~\cite{kuznetsov2017fermi}). Recently, the important results were obtained by G.~Biondini et al.~\cite{biondini2016universal,biondini2016oscillation}. They analytically described the MI of continuous spectrum waves via asymptotic analysis of the GLME.

For a long time, the role of discrete spectrum solutions in the development of SLCP was attributed to only a particular class of Peregrine-type rational solutions, which are also known as the simplest analytical model of rogue wave formation (see, e.g.,~\cite{akhmediev2009rogue,dubard2013multi}). Later V.I.~Shrira and V.V.~Geogjaev expanded this approach to the wide subclass of Kuznetsov-Ma breathers~\cite{shrira2010makes}. Recently, V.E.~Zakharov and A.A.~Gelash found that the so-called superregular breathers may be hidden in SLCP at the moment of their pairwise collisions. Based on this observation, they suggested a new scenario of MI development, which describes the formation of moving interacting breathers~\cite{zakharov2013nonlinear,gelash2014superregular}. 

Here we present analytical theory of breather collisions and numerical study of random SLCP, that show the key role of Kuznetsov-Ma and superregular breathers in the formation of rogue waves. Our theory covers all known by now scenarios of the presence of breathers in SLCP.

\subsection{Breather solutions of the NLSE}
The eigenvalue problem for ZH system is written as:
\begin{equation}\label{ZH system1}
\widehat{L}\mathbf{\Phi} = \lambda \mathbf{\Phi}\,,
\,\,\,\,\,\,\,\,\,\,\,\,\,\,\,\,\,
\widehat{L} = i\begin{pmatrix}\ 1 & 0 \\ 0 & -1 \end{pmatrix}\frac{\pt}{\pt x} - i\begin{pmatrix}\ 0 & \psi \\ \psi^* & 0 \end{pmatrix}\,.
\end{equation}
Here, $\mathbf{\Phi}(x,t,\lambda)$ is the eigenfunction, and $\lambda$ is the spectral parameter. The general $N$-breather solution of the NLSE~(\ref{NLSE}) corresponds to $N$ discrete eigenvalues $\lambda_n$ on the upper half of the $\lambda$-plane and can be written in the following form~\cite{zakharov2013nonlinear,gelash2014superregular}:

\begin{small}
\begin{eqnarray}\label{NBsolution}
\psi_N = A +
2\cdot\mydet
\left(\begin{array}{cc}
        0 & \begin{array}{ccc}
              q_{1,\beta} & \cdots & q_{n,\beta}
            \end{array}
         \\
        \begin{array}{c}
          q^*_{1,\alpha} \\
          \vdots \\
          q^*_{n,\alpha}
        \end{array}
         &  \begin{array}{c}
              \hat{M}^{T}
            \end{array}
      \end{array}\right)
(\mydet \hat{M})^{-1}.
\end{eqnarray}
\end{small}

Here the matrix $\hat{M}_{nm}=i(\mathbf{q}_{n}\cdot \mathbf{q}^*_{m})/(\lambda_{n} - \lambda^*_m)$, and the two-component vectors $\mathbf{q}_n = (q_{n,1},q_{n,2})$ are given by:
\begin{eqnarray}
q_{n,1} = \E^{-\phi_n} - \frac{\I A \E^{\phi_n}}{\lambda_n+\zeta_n}\,,
\,\,\,\,
q_{n,2} =  -\frac{\I A \E^{-\phi_n}}{\lambda_n+\zeta_n} + \E^{\phi_n}\,,
\\\nonumber
\phi_n = -i \zeta_n x - \myIm [\zeta_n] x_{0,n} - \I \lambda_n \zeta_n t - \I\theta_n/2 \,,
\end{eqnarray}
where the functions $\zeta_n = \sqrt{\lambda_n^2 + A^2 }$ define the branchcut of the spectral parameter plane on the interval $[-iA, iA]$. Each $n$-th breather is characterised by four real parameters: $\myRe [\lambda_n]$, $\myIm [\lambda_n]$, $x_{0,n}$ and $\theta_n$. The complex eigenvalue $\lambda_n$ is responsible for the breather amplitude, velocity and period of oscillations, while $x_{0,n}$ and $\theta_n$ describe the breather position in space and complex \textit{internal} phase (i.e. $\theta_n$ cannot be written as common multiplier $\E^{\I \theta_n}$ for a single breather).

The single-breather solution is defined by parameters $\lambda, \, x_0, \, \theta$, and by one vector $\mathbf{q} = (q_1,q_2)$:
\begin{eqnarray}\label{1Bsolution}
\psi_1 = \E^{-\I \theta_c} \biggl(A + 2i( \lambda + \lambda^*) \frac{q_1^* q_2}{|q_1|^2 + |q_2|^2} \biggr)\,.
\end{eqnarray}
Here, $\theta_c$ is the additional phase parameter, which is the general phase of the condensate and usually is taken to be zero. However, in this work, $\theta_c$ is very useful in the description of breather interactions -- see the text below.

\begin{figure}[h]
\centering
\includegraphics[width=3.4in]{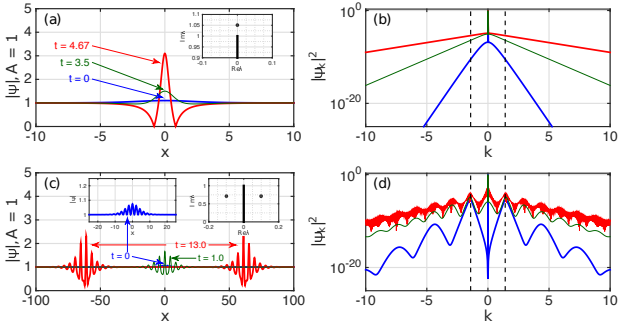}
\caption{\label{increment_peak_kspace}
Perturbation development driven by Kuznetsov-Ma breather (a,b) and the simplest one-pair superregular breather (c,d). (a,c): amplitude profiles; (b,d): evolution of the corresponding Fourier spectra (dotted lines correspond to $k_{max}$). The magnification of the initial condensate perturbation and the eigenvalues of breathers are shown in the insets.}
\end{figure}

The Kuznetsov-Ma breather is particular case of the general solution~(\ref{1Bsolution}) at $ \myRe [\lambda]=0,\,\myIm [\lambda]>A$. It is standing oscillating object localized in space (Fig.~\ref{increment_peak_kspace}). At the moment of minimum amplitude Kuznetsov-Ma breather can generate SLCP. The amplitude of the perturbation is controlled by the distance $\ae$ ($\lambda = iA+i\ae$) from the eigenvalue to the branch point, while its characteristic width is proportional to $1/\ae$ (the degenerate limit at $\ae\ra 0$ corresponds to Peregrine solution). Thus, the Kuznetsov-Ma scenario of SLCP development can be implemented when $\ae \ll 1$.

As found in~\cite{zakharov2013nonlinear}, for certain parameters, $\widetilde{N}$ pairs of breathers can describe the development of MI from SLCP. These special types of exact solutions are called $\widetilde{N}$-pair superregular breathers. In the simplest case, the superregular scenario of MI development is described by one pair of breathers with symmetrically located (about the branchcut) eigenvalues and the following space-phase synchronization:
\begin{eqnarray}\label{symmetric_params}
\lambda_{1,2} = i\mu \pm \varepsilon,  \quad \theta_{1,2}=\pi/2, \quad x_{0,1} = x_{0,2}, \quad \mu < A.
\end{eqnarray}
The distance from the eigenvalues to the branchcut $\varepsilon \ll 1$ characterises the amplitude of initial perturbation and $1/\varepsilon$ is responsible for the perturbation width. In Fig.~\ref{increment_peak_kspace}, we present an example of a symmetric one-pair superregular solution~(\ref{symmetric_params}) with $\mu=1/\sqrt{2}$, which corresponds to the maximum of the MI growth rate~(\ref{MI increment}) at $k_{max}=\sqrt{2}$. In general case, superregular breather eigenvalues can be located slightly asymmetrically and phases of the breathers can approximately satisfy the condition $\theta_1 +\theta_2 = \pi$, see the detailed discussion in~\cite{gelash2014superregular} and also Appendix section.

\subsection{Analytical theory of rogue waves formation}

\begin{figure}
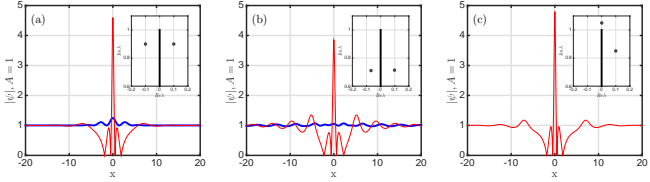

\centering
\includegraphics[width=1.05in]{Fig2a.eps}
\,
\includegraphics[width=1.05in]{Fig2b.eps}
\,
\includegraphics[width=1.05in]{Fig2c.eps}
\caption{\label{Fig3}
Amplitude profiles of two-breather collisions. (a,b): bold blue lines correspond to superregular synchronization ($\theta_1=\theta_2=\pi/2$); thin red lines show the rogue wave formation ($\theta_1=\theta_2=\pi$). (a) corresponds to completely symmetric and (b) to slightly asymmetric breather eigenvalues. (c): collision of Kuznetsov-Ma breather with one breather from superregular pair ($\theta_1=\theta_2=\pi$). The corresponding Fourier spectra for these and subsequent wavefields are presented in Appendix section.}
\end{figure}

Collisions of NLSE breathers with particular phases can produce typical rogue waves~\cite{pelinovsky2008book,akhmediev2009excite,akhmediev2009extreme}. The formation of SLCP by superregular solutions is the back side of this process. Indeed, depending on the sum of the breather phases, $\theta_1+\theta_2$, the amplitude profile at the moment of collision can vary from a small perturbation to a rogue wave of extreme amplitude, which we discussed in~\cite{kibler2015superregular} and illustrate in Fig.~\ref{Fig3},a,b. The rogue waves formation is also possible in the case of interaction of superregular and Kuznetsov-Ma breathers, that we show in Fig.~\ref{Fig3},c.

To analyse the multi-breather interactions, we need exact expressions describing the space-phase shifts that breathers acquire after mutual collisions. Such formulas have been known for soliton interactions  since the paper~\cite{zakharov1972exact}. However, the derivation of the space-phase shifts for breathers was skipped in the work devoted to the IST of the NLSE on the condensate background~\cite{kuznetsov1977solitons,ma1979perturbed,kawata1978exact}. To the best of our knowledge, these expressions have not been reported.

We solve the space-phase shifts problem using the standard technique: asymptotic analysis of the two-breather solution (in the general case of arbitrary eigenvalues $\lambda_1$ and $\lambda_2$), which can be obtained from~(\ref{NBsolution}) at $N=2$. Considering the asymptotic states of separate breathers at $t\ra \pm \infty$, we find the following expressions that describe the space shift $\Delta x_{0,2}$ and phase shifts $\Delta \theta_2, \Delta \theta_{c,2}$ that breather $2$ acquires after collision with the breather $1$:
\begin{eqnarray}\label{shift_exp}
\Delta x_{0,21} = ln \bigl[ (s_1-s_3)/(s_2-s_4) \bigr]/2 Im[\zeta_2]\,,
\\\nonumber
\Delta \theta_{21} = 2\myArg [i (p_1 + p_2)]\,, \Delta \theta_{c,21} = -4 \myArg [\lambda_1+\zeta_1]\,.
\end{eqnarray}
Here, the coefficients $s_1, s_2, s_3, s_4, p_1$, and $p_2$ have a cumbersome dependence on the eigenvalues $\lambda_1$ and $\lambda_2$:
\begin{small}
\begin{eqnarray}\nonumber
s_1 = (A^4 + |\lambda_1 + \zeta_1|^2 \cdot |\lambda_2 + \zeta_2|^2) \cdot |\lambda_1 - \lambda^*_2|^2 +
\\\nonumber
+ A^2 (|\lambda_1 + \zeta_1|^2 + |\lambda_2 + \zeta_2|^2) \cdot |\lambda_1 - \lambda_2|^2\,,
\\\nonumber
s_2 =  A^2 (|\lambda_1 + \zeta_1|^2 + |\lambda_2 + \zeta_2|^2) \cdot |\lambda_1 - \lambda^*_2|^2 +
\\\nonumber
+ (A^4 + |\lambda_1 + \zeta_1|^2 \cdot |\lambda_2 + \zeta_2|^2) \cdot |\lambda_1 - \lambda_2|^2\,,
\\\nonumber
s_3 = A^2 (\lambda_1 - \lambda^*_1)(\lambda_2 - \lambda^*_2) \cdot [(\lambda_1 + \zeta_1)(\lambda_2 + \zeta_2) +
\\\nonumber
+ (\lambda^*_1 + \zeta^*_1)(\lambda^*_2 + \zeta^*_2)]\,,
\\\nonumber
s_4 = -A^2  (\lambda_1 - \lambda^*_1)(\lambda_2 - \lambda^*_2) \cdot [(\lambda_1 + \zeta_1)(\lambda^*_2 + \zeta^*_2) +
\\\nonumber
+ (\lambda^*_1 + \zeta^*_1)(\lambda_2 + \zeta_2)]\,;
\\\nonumber
p_1 = \{A^2 (\lambda_2 + \zeta_2 - \lambda^*_1 - \zeta^*_1) -  |\lambda_1 + \zeta_1|^2 (\lambda^*_2 + \zeta^*_2) +
\\\nonumber
+ |\lambda_2 + \zeta_2|^2 (\lambda_1 + \zeta_1)\}/\{|\lambda_1 - \lambda^*_2|^2\}\,,
\\\nonumber
p_2 = \frac{(A^2 + |\lambda_1 + \zeta_1|^2) (\lambda_2 + \zeta_2 - \lambda^*_2 - \zeta^*_2)}{(\lambda_1 - \lambda^*_1)(\lambda_2 - \lambda^*_2)}\,.
\end{eqnarray}
\end{small}
The additional general phase $\theta_c$ (defined in~(\ref{1Bsolution})) describes how the first breather reverses the condensate phase after itself. One can check that as $A\ra 0$, expressions~(\ref{shift_exp}) become well-known soliton space-phase shifts, so the two phases $\theta_2$ and $\theta_{c,2}$ form a single soliton phase.

\begin{figure}
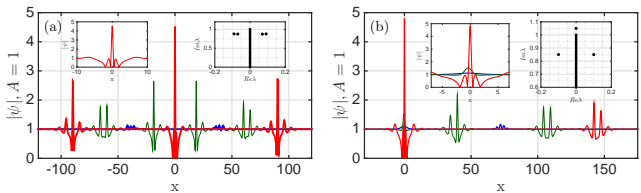

\centering
\includegraphics[width=1.6in]{Fig3a.eps}
\,
\includegraphics[width=1.6in]{Fig3b.eps}
\caption{\label{Fig4}
New analytical scenario of rogue wave formation from the condensate locally perturbed by: (a) two pairs of superregular breathers and (b) one Kuznetsov-Ma and one pair of superregular breathers. The blue lines show the initial SLCP, the thin green lines -- intermediate stage of the process and the red lines demonstrate the rogue waves formation produced by breathers collisions at the final stage.}
\end{figure}

Now, we suggest analytical description of rogue waves formation from SLCP driven by superregular and Kuznetsov-Ma breathers. Our first scenario demands at least two initial superregular perturbations, which develop into four breathers. Then, two of the breathers collide leading to producing of the rogue wave. The second scenario can be implemented by one superregular and one Kuznetsov-Ma perturbation. In this case the rogue wave is produced by collision of a moving breather generated from the superregular perturbation with standing Kuznetsov-Ma breather. The breather phases required  to generate SLCP or rogue wave in pairwise interaction are known (Fig.~\ref{Fig3}). However, in case of more than two breathers, all phases are shifted by mutual breather interactions. We account this shifts analytically and present the new scenarios of rogue wave formation in Fig.~\ref{Fig4}. The details of calculation of appropriate space-phase shift parameters for all breathers are given in the Appendix section.

\subsection{Randomly perturbed condensate}
\begin{figure}[h]
\centering
\includegraphics[width=3.4in]{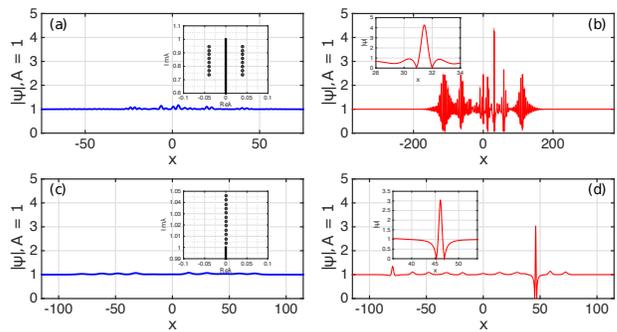}
\caption{\label{Fig5}
Rogue wave formation from the condensate locally perturbed by $8$ pairs of randomly distributed superregular breathers (a,b); $8$ randomly distributed Kuznetsov-Ma breathers (c,d). (a,c) show the initial SLCP, (b,d) correspond to the moments of maximum amplitude generation.}
\end{figure}

The suggested scenarios of rogue wave formation can be observed not only for manually synchronized breathers but also for the random distribution of condensate perturbations. In Fig.~\ref{Fig5},a, we present an example of $16$-breather MI development. At the initial moment of time, the condensate is locally perturbed by $8$ superregular pairs randomly distributed in space. In Fig.~\ref{Fig5},b we show the most significant rogue wave observation over $100$ random initial condition runs. In Fig.~\ref{Fig5},c,d we demonstrate SLCP formed by $8$ Kuznetsov-Ma breathers with synchronised phases and random positions. Now the rogue waves are generated by standing oscillations of breathers and thus have a smaller amplitude, than in the case of breather collisions. Here we synchronise phases of breathers only to produce initial SLCP, while formation of rogue waves is random. In this case ensemble of symmetric superregular breathers can be synchronised without using formulas~(\ref{shift_exp}), since the phase shift to one breather in a pair (from the breathers around) is compensated by the shift obtained by the second breather in the pair. Meanwhile synchronisation of an ensemble of Kuznetsov-Ma breathers always demands to account all mutual phase shifts. Thus, Fig.~\ref{Fig5}(c,d) explicitly demonstrate scenario of rogue waves formation from random ensembles of Kuznetsov-Ma breathers for the first time. Note, that superregular scenario can describe the development of the most unstable Fourier modes, while Kuznetsov-Ma perturbations always develop slowly -- see the Fig.~\ref{increment_peak_kspace} and the Fourier spectra in the Appendix section.

The simulation of discrete spectrum solutions by exact IST formulas is an ill-conditioned problem at a large $N$, that we recently discussed for the case of solitons in~\cite{gelash2016uniform,frumin2017new}. Here, for breathers we use the Zakharov-Shabat dressing method (an alternative formulation of the solution~(\ref{NBsolution}) -- see~\cite{zakharov2013nonlinear,gelash2014superregular} for details) supplemented with $100$-digits precession arithmetic to mitigate the numerical instability and to generate accurate solutions.

The final question is what type of breathers are contained in arbitrary-type perturbations. We use $m \sim 10$ harmonics with arbitrary phases $\varphi_j$ belonging to the unstable part of the MI growth rate~(\ref{MI increment}) and modulate them with different functions $f(x,\sigma)$ of the spatial width $\sigma$ to obtain general type condensate perturbation:

\begin{small}
\begin{eqnarray}\label{myperturbation}
\psi(x) = A + \epsilon\rho f(x,\sigma)\sum^{m/2-1}_{j=-m/2} \sin((k_0 + j\Delta k) x + \varphi_j).
\end{eqnarray}
\end{small}

Here $\epsilon$ characterises the perturbation amplitude and $\rho$ is a complex constant. Then, we solve the eigenvalue problem~(\ref{ZH system1}) using the standard Fourier collocation method~\cite{yang2010nonlinear} with high numerical accuracy. We study the initial perturbation at a large numerical interval of $1024\cdot\pi$ and use 16384 discretization points (the latter corresponds to the maximum performance of 192 GB RAM). This process enables us to avoid the effect of periodicity (see e.g.~\cite{ablowitz1996computational,grinevich2017finite,randoux2016inverse}) and unambiguously identify discrete spectrum eigenvalues.
 
\begin{figure}[h]
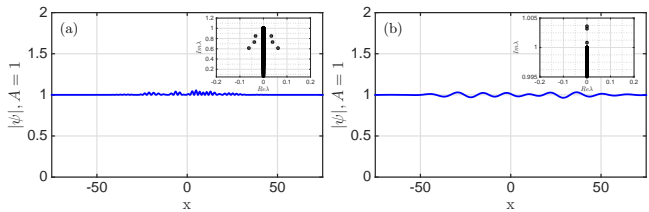

\centering
\includegraphics[width=1.65in]{Fig5a.eps}
\includegraphics[width=1.65in]{Fig5b.eps}
\caption{\label{Fig6} Eigenvalues of arbitrary-type SLCP, generated via~(\ref{myperturbation}) with $f(x)=f_2(x)$. (a) imaginary perturbation with $m=8$, $k_0 = \sqrt{2}, \Delta k = 0.1, \sigma=30$, $\epsilon = 10^{-1}$. (b) real perturbation with $m=8, k_0 = 0.4, \Delta k = 0.025,\sigma=40, \epsilon = 2\cdot 10^{-2}$.}
\end{figure}

Symmetric superregular breathers~(\ref{symmetric_params}) generate pure imaginary condensate perturbation to the real-valued condensate. This can be easily proved by considering a two-breather solutions with parameters~(\ref{symmetric_params}) at $t=0$. Thus, we first study pure imaginary arbitrary-type condensate perturbations, i.e., the case $\rho=\I$ in~(\ref{myperturbation}). For several random runs with modulation functions $f_1= \mysech(\sigma x)$ and $f_2= \exp(-x^2/\sigma^2)$, we clearly observe different distributions of symmetric eigenvalues near the branchcut -- an example is presented in Fig.~\ref{Fig6},a. Then we study pure real perturbations (Kuznetsov-Ma breather generate real-valued perturbation). Now for $\rho=1$ we find eigenvalues on the real axes very close to the branch point -- see Fig.~\ref{Fig6},b.

\subsection{Discussion and Conclusion}
The eigenvalue spectrum does not reveal the impact of the continuous spectrum waves (we also need to study the \textit{reflection coefficient}, see, e.g.,~\cite{kuznetsov1977solitons,ma1979perturbed}). The numerical simulation of the evolution of the perturbations presented in Fig.~\ref{Fig6} shows complicated wave patters that are apparently driven by nonlinear interaction between discrete and continuous spectrum solutions. Another important task is to study the combinations of imaginary and real perturbations or perturbations based on broadband random noise. For the latter case, B.~Kibler et al. found strong signatures of superregular breathers at the intermediate stage of MI development~\cite{wabnitz2017book}. All these questions demand separate consideration.

Another fundamental problem is the development of MI from spatially periodic perturbations (see e.g.~\cite{osborne2010nonlinear}). Recently, D.S.~Agafontsev and V.E.~Zakharov have found that in this case, MI driven by small-amplitude perturbations ($\sim 10^{-5} A$) leads to the formation of Gaussian wavefield statistic. Later, J.M.~Soto-Crespo, N.~Devine and N.~Akhmediev have demonstrated that in the case of higher condensate disturbances, the tails of the probability density function of wave amplitudes increases~\cite{soto2016integrable} (see also the works~\cite{walczak2015optical,suret2016single,soto2016integrable} where the non-Gaussian statistics was observed for the development of strongly fluctuating initial conditions). They concluded that only initial perturbations of significant amplitude can contain spatially localized breathers, that is consistent with the theory suggested here (the minimal amplitude of SLCP generated by breathers is determined by the perturbation width, see the second paragraph). Meanwhile S.~Randoux, P.~Suret and G.~El have suggested that the key role in the development of periodic perturbations should be attributed to the \textit{finite-band} solutions of the NLSE~\cite{randoux2016inverse} (see also~\cite{bertola2016rogue}). The problems of localized and periodic condensate perturbations are complimentary and both have fundamental importance. The understanding of the link between the localized and periodic IST description of MI is an important task. The obtained pictures of eigenvalues for real and imaginary SLCP correlate with results presented earlier for the periodic perturbations~\cite{ablowitz1996computational}, that can be a starting point for such study.

The found space-phase shifts (\ref{shift_exp}) are critically important to describe interactions of NLSE breathers when their number $N\geqslant 3$. They can be used for further studies, among which the experimental realization of complicated multi-breather dynamics is of special interest. Similar breathers describe the dynamic of unstable backgrounds in different integrable models (see e.g.~\cite{breizman1980dynamics,ankiewicz2010rogue,kedziora2014rogue,liu2017superregular}), that allows to generalise our results.

\section{APPENDIX}

\subsection{Uniformization of spectral parameter $\lambda$}

In our previous works~[15,22,23,37] we use Joukowsky transform (uniformization) of standard spectral parameter $\lambda$ to simplify analysis of $N$-breather solutions:
\begin{equation}\label{uniformization}
\lambda = -\frac{iA}{2} \biggl(\xi+\frac{1}{\xi}\biggr), \quad\quad\quad \xi = re^{\I\alpha} = e^{z + \I\alpha}\,.
\end{equation}
In such variables the two-sheeted Riemann $\lambda$-surface transforms into the one-sheeted uniformized $\xi$-plane. The line of the branchcut $[-iA,iA]$ transforms into a circle of unit radius, while the Riemann sheets become the outer and inner pars of the circle. The purpose of this supplemental paragraph is to illustrate the relation between representations of the breather solutions in $\lambda$ and $\xi$ variables.

In the work [22] we found, that asymptotics of the general one-breather solution (Eq.(6) with $\theta_c=0$) can be simply expressed using uniformized variables (\ref{uniformization}):
\begin{equation}
\psi_1 \ra A \exp(\pm 2i\alpha)\,, \quad\quad\quad |x|\ra\infty \,.
\end{equation}
The latter means that single breather changes the phase of the condensate before and after itself on $4\alpha$. Now the requirements for breathers which can generate SLCP can be easily obtained.
Indeed, the condensate complex phase at infinity should be constant with time for any localized perturbation. Thus, the breather solutions capable to describe the growth of SLCP should have equal condensate phases at $|x|\ra\infty$, or at least have difference in the phases comparable with the relative amplitude of the initial perturbation.

Among one-breather solutions of the NLSE, only Kuznetsov-Ma and Peregrine breathers have equal condensate phases at infinity since they correspond to the case $\alpha = 0$. For a couple of breathers, the total change of phase before and after them is zero when
\begin{equation}\label{alphacond}
\alpha_1 = -\alpha_{2} = \alpha\,.
\end{equation}
As was found in the work~[22], at certain phase synchronisation (close to $\theta_1+\theta_2=\pi$) and when the breather eigenvalues located near the branchcut ($r-1=\tilde{\varepsilon}\ll 1$), such two-breather (superregular) solution generate SLCP at an initial moment of time.

For symmetric one-pair superregular solution (Eq.(7)) the initial (linear) stage of perturbation growth is described by the following expression obtained in~[23]:
\begin{eqnarray}\label{SRapproximation}
\delta\psi \approx
\I\tilde{\varepsilon} A
\frac{\cosh(\I\alpha-2\gamma t)\cos (2\eta x)}{\cosh((2A\tilde{\varepsilon}\cos\alpha)x)},
\\\nonumber
\eta=A\cosh z\sin\alpha, \quad\quad \gamma= -A^2\cosh 2z\sin 2\alpha/2.
\end{eqnarray}
Here $\delta\psi$ is the SLCP: $\psi = A + \delta\psi$. The maximum value of the growth rate (defined by $2\gamma$) for the expression~(\ref{SRapproximation}) is achieved at $\alpha = \pi/4$. In this case superregular breather corresponds to the maximum of the MI growth rate~(2). Indeed, when $\alpha = \pi/4$ ($Im[\lambda] = 1/\sqrt{2}$), maximum of the most significant Fourier spectrum sideband mode is located at $k_{max} = \pm \sqrt{2}$ (with accuracy $\sim\tilde{\varepsilon}$) -- see the Fig.1. Note, that the similar result for the Akhmediev breather is well known -- see the reference~[43].

\begin{figure}[h]
\centering
\includegraphics[width=1.65in]{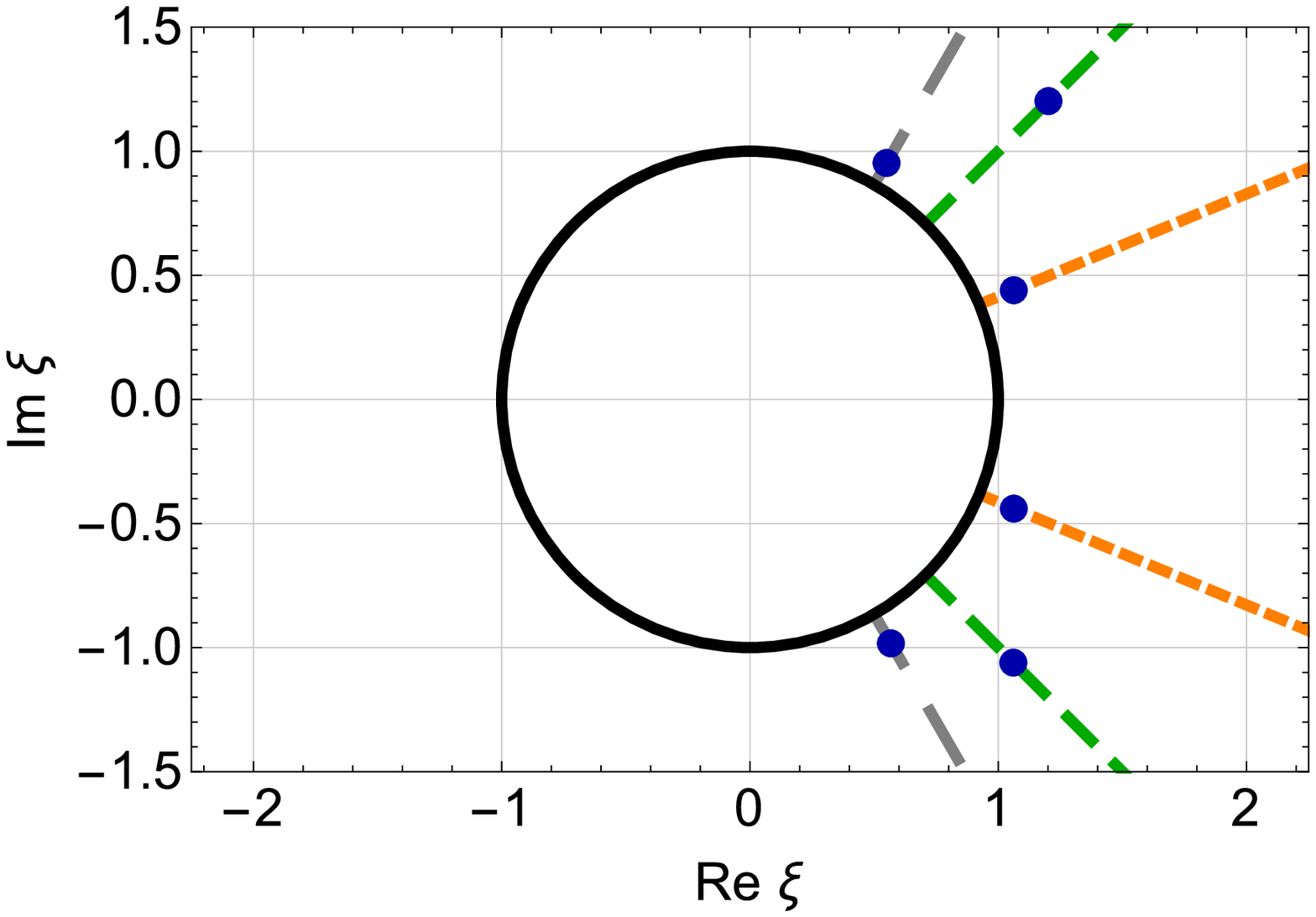}
\includegraphics[width=1.65in]{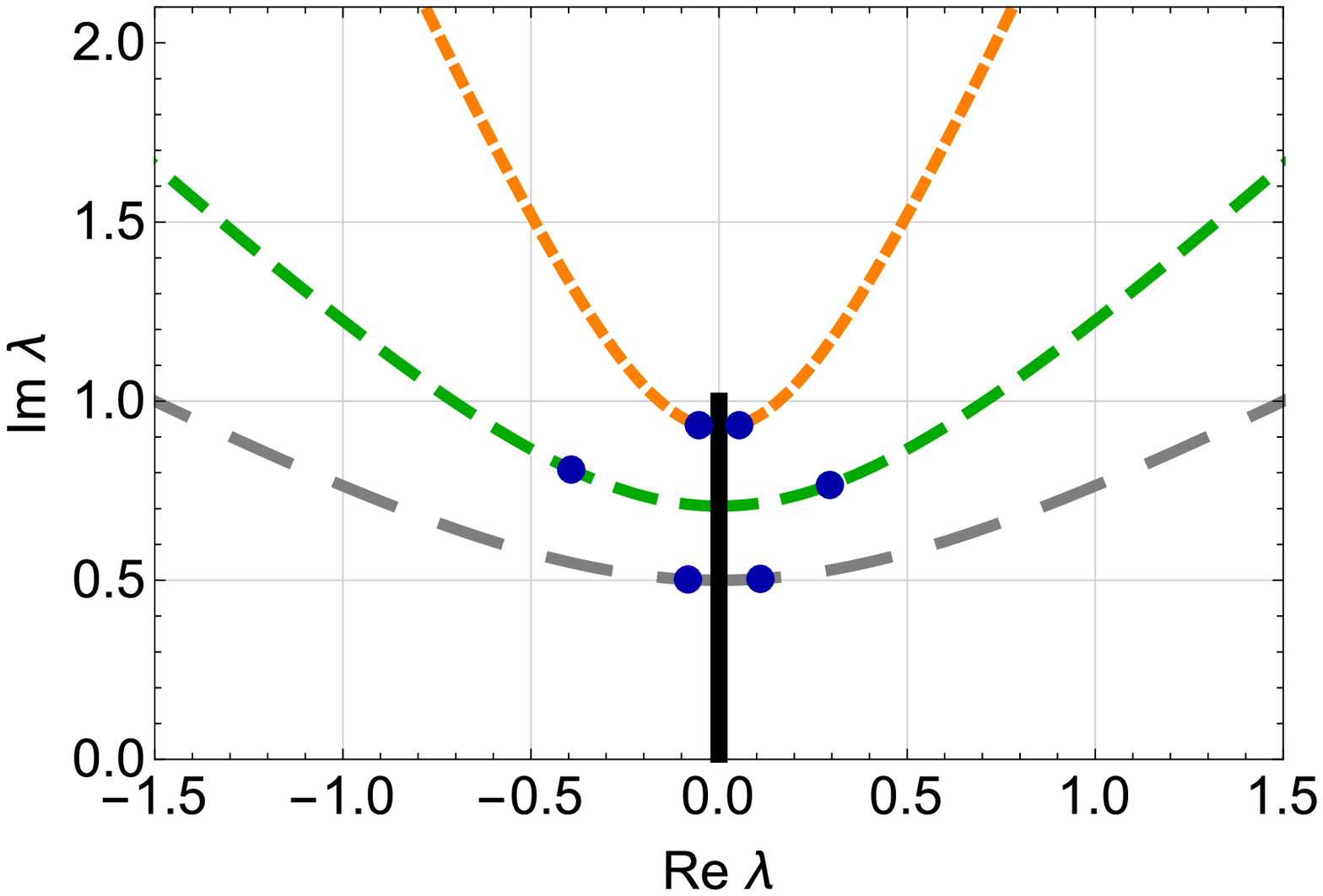}
\caption{\label{suppFig1}
Comparison of $\xi$ and $\lambda$ parametrizations of spectral parameter. The branch cut and its Joukowsky mapping~(\ref{uniformization}) are drawn by black solid lines. The pairs of breather eigenvalues (marked by blue points) lie on the rays~(\ref{alphacond}) (dashed lines, left picture) and on the parametric curves~(\ref{parametric}) (dashed lines, right picture).}
\end{figure}

The relation~(\ref{alphacond}) defines the family of the following parametric curves 
\begin{eqnarray}\label{parametric}
\myRe [\lambda]  = \pm\frac{A\sin\alpha}{2}\biggl(r - \frac{1}{r}\biggr)\,,
\,\,
\myIm [\lambda]  = \frac{A\cos\alpha}{2}\biggl(r + \frac{1}{r}\biggr)\,,
\end{eqnarray}
that is illustrated here in Fig.~\ref{suppFig1}.  The SLCP can be generated only by eigenvalues located on the grey (long dashed) and orange (short dashed) lines since the distance between the eigenvalues and the branchcut defines the amplitude of the perturbation.

\subsection{Breather collisions sinchronization}

Then we present details of synchronization of breather SLCP development presented by Fig.~3 and Fig.~4. First we introduce the breather group and phase velocities, that can be obtained from expressions (5):
\begin{equation}
V_{gr} = - \frac{\myRe[\lambda_n \cdot \zeta_n]}{\myRe[\zeta_n]}\,, \,\,\,\,\,\,\,\,\,\,\,\,\,\,\, V_{ph} = \myRe[\lambda_n \cdot \zeta_n]\,.
\end{equation}

For the four-breather scenario presented in Fig.~3a, we choose the initial moment of time $t_0=0$ and the moment of rogue wave formation $t_{max} = 15$ (with maximum amplitude at $x=0$). The problem is to find initial space shifts and phases for each breather. Let us denote breathers in the left superregular pair (see Fig.~3a) by indexes $1$ (the breather moving to the left) and $2$ (the breather moving to the right) and breathers in the right pair by indexes $3$ (the breather moving to the left) and $4$ (the breather moving to the right). The breathers $2$ and $3$ have complex conjugated eigenvalues, so they differ only in the direction of the group velocity: $V_{gr_2} = - V_{gr_3} = \widetilde{V}_{gr}$. Before the moment $t_0$, the breather $2$ was affected by collisions with the breathers $1$ and $4$, so the total space shift for the breather $2$ can be found as:
\begin{equation}
x_{0,2} = -\widetilde{V}_{gr} \cdot t_{max} - \Delta x_{0,12} + \Delta x_{0,42}\,.
\end{equation}
Here $\Delta x_{0,12}$ and $ \Delta x_{0,42}$ are calculated via space-shift formula (9). To obtain further synchronisation inside the left superregular pair, we choose the same space shift for the breather $1$: $x_{0,1}=x_{0,2}$. By analogy we obtain the following space shifts for the breathers $3$ and $4$:
\begin{equation}
x_{0,3} = x_{0,4} = \widetilde{V}_{gr} \cdot t_{max} - \Delta x_{0,13} + \Delta x_{0,43}\,.
\end{equation}
Synchronization of phases we start from the breathers $2$ and $3$. According to the Fig.~2 the phases of both breathers should be equal to $\pi$, but we have to account the phase changing with time $t_{max}$ and phase shifts acquired after collisions with breathers $1$ and $4$. Denoting the phase velocity for the second and third breather as $V_{ph_2} = V_{ph_3} = \widetilde{V}_{ph}$ we obtain:
\begin{equation}
\theta_2 = \theta_3  = \pi + \widetilde{V}_{ph}\cdot t_{max} - \Delta \theta_{12} + \Delta \theta_{42}\,.
\end{equation}
Here $\Delta \theta_{12}$ and $ \Delta \theta_{42}$ are calculated via phase-shift formula (8). Now the formation of the rogue wave at $t_{max}$ is obtained and we only need to synchronise formation of small condensate superregular perturbations at $t_0$. As we discussed in the main text of the paper, superregular synchronization appears at $\theta_1 + \theta_2 = \pi$ (see again the Fig.~2). The breather $1$ was affected by collisions with the breathers $3$ and $4$, while the breather $4$ collided with breathers $3$ and $1$. Thus, the required phase shifts can be found as:
\begin{eqnarray}
\theta_1 = \pi - \theta_2 - \Delta\theta_{13} - \Delta\theta_{14}\,,
\\\nonumber
\theta_4 = \pi - \theta_3 + \Delta\theta_{43} - \Delta\theta_{14}\,.
\end{eqnarray}

Now we discuss synchronization of Kuznetsov-Ma breather scenario of SLCP development, presented by Fig.~4c,d. Our task is to find phases $\theta_m$ for all $M$ breathers which correspond to minimum amplitude at $t=0$.
A single Kuznetsov-Ma breather has minimum amplitude at $t=0$ when its phase $\theta$ is equal to zero (like in the Fig.~1a). Thus, for each $m$-th Kuznetsov-Ma breather we should set zero phase and add phase correction from each of the neighbouring breathers:
\begin{eqnarray}
\theta_m = \sum_{j=1,j\ne m}^{M} \Delta\theta_{mj}
\end{eqnarray}
Where $\Delta\theta_{mj}$ is calculated using formulas (8).

The combined scenario for superregular and Kuznetsov-Ma breathers presented in Fig.~3b can be obtained in similar to the discussed above way, thus we skip its details here.
\subsection{Fourier spectra for SLCP development}

\begin{figure}
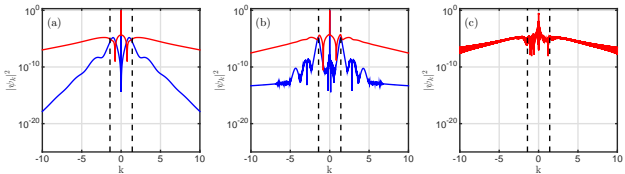

\centering
\includegraphics[width=1.05in]{suppFig2a.eps}
\includegraphics[width=1.05in]{suppFig2b.eps}
\includegraphics[width=1.05in]{suppFig2c.eps}
\caption{\label{suppFig2} Fourier spectra for different two-breather collisions, presented in Fig.~2.}
\end{figure}

\begin{figure}
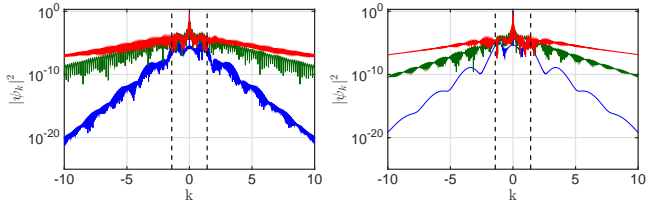

\centering
\includegraphics[width=1.65in]{suppFig3a.eps}
\includegraphics[width=1.65in]{suppFig3b.eps}
\caption{\label{suppFig3} Fourier spectra for analytical scenarios of rogue wave formation from the condensate locally perturbed by superregular (left) and one Kuznetsov-Ma and one superregular (right) breathers presented in Fig.~3.}
\end{figure}

\begin{figure}
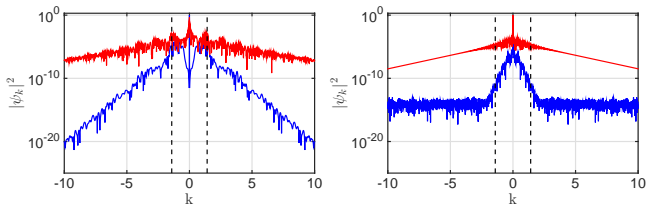

\centering
\includegraphics[width=1.65in]{suppFig4a.eps}
\includegraphics[width=1.65in]{suppFig4b.eps}
\caption{\label{suppFig4} Fourier spectra for scenarios of rogue wave formation from the condensate locally perturbed by random distributions of superregular (left) and Kuznetsov-Ma (right) breathers presented in Fig.~4.}
\end{figure}

Here in Fig.~\ref{suppFig2},~\ref{suppFig3},~\ref{suppFig4} we present Fourier spectra for all studied in the main text of the paper, scenarios of SLCP development. In general, they are in accordance with Fourier spectra of their elementary building blocks: one-breather Kuznetsov-Ma solution and one-pair superregular breather solution, which are presented in Fig.~1. Namely, the development of SLCP driven by breathers leads to the broadens of the Fourier spectrum. Interactions of several breathers make the Fourier spectrum more complicated. In all cases we can see that superregular breathers with $\mu\approx 1/\sqrt{2}$ are responsible for the fastest perturbation growth.

\subsection{Acknowledgments}
The author is thankful to 1) Prof. V.E.~Zakharov for providing the idea to study imaginary and real condensate perturbations separately; 2) Prof.~E.A. Kuznetsov for the helpful discussions of continuous spectrum waves; 3) Dr. D.S.~Agafontsev for the helpful comments with regard to the Fourier spectrum of superregular breathers and discussion of possible applications of the presented results in the interpretation of numerical simulations with periodically perturbed condensate; and 4) Dr. B. Kibler for the helpful discussions of the broadband random noise SLCP. Numerical code and simulations were performed at the Novosibirsk Supercomputer Center (NSU). The first part of the work was performed with support from the Russian Science Foundation (Grant No. 14-22-00174). The study presented in the last part "Randomly perturbed condensate" was supported by the RFBR (Grants No. 16-31-60086 mol\_a\_dk and 17-41-543347 r\_mol\_a).

\bibliographystyle{apsrev4-1}

\end{document}